\documentclass[a4paper]{article}
\usepackage{graphics}
\usepackage{graphpap}
\usepackage{color}
\usepackage[only,bbm10]{rawfonts}
\newfont{\Bb}{bbm10}
\newcommand{\be}{\begin{equation}}
\newcommand{\ee}{\end{equation}}
\newcommand{\Q}{\hbox{\Bb Q}}

\newcommand{\R}{\hbox{\Bb R}}
\newcommand{\M}{\hbox{\Bb M}}
\newcommand{\pa}{\partial}

\newcommand{\vep}{\varepsilon}

\newcommand{\al}{\alpha}
\newcommand{\bt}{\beta}
\renewcommand{\t}{\tau}
\newcommand{\dq}{\dot{q}}
\newcommand{\px}{{x^{\prime}}}
\newcommand{\pt}{\theta^\prime}
\newcommand{\cL}{{\mathcal L}}
\newcommand{\cP}{{\mathcal P}}
\newcommand{\cR}{{\mathcal R}}
\newcommand{\cW}{{\mathcal W}}
\newcommand{\bl}{\left}
\newcommand{\br}{\right}

\newcommand{\te}{\theta}
\newcommand{\la}{\label}

\begin{document}
\bibliographystyle{plain}
\begin{titlepage}
\begin{center}
{\Large\bf SOME THEOREMS RELATED TO THE}\\ 
\vspace{4pt}
{\Large\bf JACOBI VARIATIONAL PRINCIPLE OF}\\ 
\vspace{4pt}
{\Large\bf ANALYTICAL DYNAMICS}\raisebox{.6ex}{$^{*\dag}$}\\
\vspace{2cm}
{Stanis{\l}aw L.~Ba\.za\'nski}\\
\vspace{.3cm}
{\small{\it Dept.~of Physics, Institute of Theoretical Physics, Warsaw 
University, Poland}}\\
{\small{\it E-mail: bazanski@fuw.edu.pl}}
\end{center}
\vspace{2cm}
\begin{abstract}
 {It is shown that there exists a commuting diagram of mappings
between dynamics of classical systems on one side and variational
principles for geodesic lines in stationary spacetimes of general
relativity on the other. The construction of the mappings is based on
classical Routh's and Jacobi's reduction procedures and on corresponding
inverse procedures which are reviewed in the paper.}
\end{abstract}
\vspace{6cm}

{\small
Subj.~-class: Mathematical Physics, Differential Geometry, General 
Relativity}

{\small
MSC-class: Primary: 70H30, 53B30. Secondary: 83C20, 53B50, 83-02.}
\vfill
\footnoterule
{\footnotesize

$^*$I am honoured that I may
dedicate this work to Jerzy Pleba\'nski, a teacher and a friend of mine
who a long time ago introduced me into the realm of variational 
principles in physics, on the occasion of his 75$^{\rm th}$ birthday.

$^\dag$This is a contribution to a volume in celebration of the
75$^{\rm th}$ birthday of Jerzy Pleba\'nski. Due to a considerable delay
of its publication, I have decided now to publish it in a preprint form.
}
\end{titlepage} 

{\noindent\bf 1.~Introduction}
\vspace{.1cm}

{\noindent Since a long time it has been well known that the number of
Newtonian differential equations of motion can be diminished by making use
of the existence of some mechanical conservation laws. In the middle of
the 19$^{\rm th}$ century a new problem of this kind was posed. If the
original set of equations of motion are the Euler-Lagrange equations of a
Lagrangian, and the dynamical system admits conservation laws that can be
used to reduce the number of these equations, is then always possible to
find a new Lagrangian such that the reduced system of differential
equations can be derived as Euler-Lagrange equations of the new
Lagrangian?}

As is known, cf.~\cite{LL}, the first solution to the problem was given in
1876 by E.~J.~Routh who showed that when the corresponding conservation
laws resulted from the occurrence of cyclic dynamical variables in the
original Lagrangian, it was the Routh function that was the Lagrangian for
the reduced system. If, however, the law responsible for the reduction of
the system is the conservation of energy, the Routh method applied
directly to the original Lagrangian does not work at all. This case
required a separate treatment that was given in 1886 in a book by
K.~G.~J.~Jacobi, where a variational principle leading to differential
equations satisfied by spatial trajectories in the configuration space of
the dynamical system was formulated under the assumption that the original
Lagrangian did not depend explicitly on the time variable. The proof of
Jacobi was based on the Maupertuis principle of least action. This fact
may be one of the sources of a terminological confusion which appears in
many contemporary text-books on analytical dynamics, where the Jacobi
principle is named Maupertuis principle, despite the fact that the latter
is, for holonomic dynamical systems and for $E\neq 0$, equivalent to the
Lagrange equations of the second kind which determine the motion of the
system, whereas the Jacobi principle determines only the orbits of the
motion.

In 1994, in \cite{BJ}, the present author together with P.~Jaranowski
have posed and solved, as they have named it, the inverse Jacobi problem:
under which conditions imposed, can one restore the original motion when
starting from a variational principle leading to orbits?

In \cite{BJ} an attempt was also made to derive {\it ab initio} the
standard Jacobi principle without making any use of the Maupertuis
principle.  During my seminar talks on the results obtained in
\cite{BJ}, I realized that the ``new'' derivation of the Jacobi
principle was rather complicated to convey it to the audience. In 2001, I
found a very simple derivation of the Jacobi principle, published later in
\cite{B}. It makes use of the Routh method applied to an in a
suitable way transformed Hamilton's action. It is so simple that it must
have been undoubtedly known to some people before, although I could not 
find any references to it. From the point of view of methodology, this
derivation is more suitable for a classroom than the traditional one,
because it solves two akin problems in the same way.

The main objective of the article is to demonstrate that there exist
mappings between dynamics of classical systems on one side and variational
principles for geodesic lines in stationary spacetimes of general
relativity on the other. The construction of the mappings is based on
classical Routh's and Jacobi's reduction procedures and on corresponding
inverse procedures which are proposed by the present author and
P.~Jaranowski in \cite{BJ} and \cite{B}.  All these procedures
are general theoretical methods that belong to analytical dynamics. A
review of them is presented in sections 2, 3, 4, and 5, mainly in order to
fix the framework which will be employed in the next sections.

Sections 6, 7, and 8 present the classical Jacobi procedure in the
working. In Sec.~6, the relation between two widely known actions for
geodesics on manifolds is interpreted in terms of the Jacobi reduction of
the ``quadratic'' action into the other one. Section 7 repeats the
elementary text-book example of the Jacobi reduction of a Newtonian,
holonomic Lagrangian into the Lagrangian describing orbits as geodesics in
the kinetic energy metric. The action for geodesic lines in a stationary
Lorentzian manifold in the coordinate time parametrization is in Sec.~8
Jacobi reduced into an action defined on the constant time hypersurface.
All the examples considered in these three sections are at the end of
Sec.~8 reinterpreted in terms of mappings of some of the dynamics into the
other ones, and the equivalence of some of the dynamics is exhibited 
there.

The inverse Jacobi procedure is applied to the action considered in the
previous section in Sec.~9. Its result is an action determining affinely
parametrized geodesics, and the metric coefficients in this action are
time independent. This fact enables one to form a composition of the Routh
reduction with the inverse Jacobi procedure performed just at the beginning
of the section. The outcome is a dynamics which is equivalent to the
Newtonian dynamics considered in Sec.~7. This enables one to continue the
discussion led at the end of the previous section and to construct two
closed loops of mappings that alternatively can be considered as a
commuting diagram of mappings between all the dynamics dealt with in this
article. Two of the branches in this diagram may be regarded as
generalizations of the correspondences between dynamics that were already
discussed in the literature, cf.~\cite{ACL} and \cite{E}, but by methods
that are rather particular, and without any reference to general
principles of analytical dynamics.

In the text which follows, an abbreviated notation is used, in accordance
with which expressions like e.g.~$(q^i, \dq^j)$ stand for sequences $(q^1,
q^2, \dots , q^n, \dq^1, \dots ,\dq^n)$ or, depending on ranges in which
the indices vary, for some other sequences of a similar type. The
summation convention is employed throughout the article.

\vspace{.4cm}
{\noindent\bf 2.~Routh's theorem}

\vspace{.1cm}
{\noindent Let  
\be
   {\cW}[q^\al] =  \int\limits_{t_1}^{t_2} \cL
      \bl (q^i(t), \dq^\bt (t),t \br )\, dt 
\la{eq:1.1} 
\ee
be an action functional describing a dynamical system in a configuration
space ${\Q}^{n+1}$. The local coordinates $q^\al$, where $\al ,\bt =
0,1,...,n$, of a point in ${\Q}^{n+1}$ are functions of time, $q^\al =
q^\al(t)$, called the motion of the system in ${\Q}^{n+1}$.  Let further
the Lagrangian ${\cL}$ be non-degenerate. The form of the action (1) was
written down in accordance with the assumption that ${\pa \cL } / \pa 
q^0 = 0$, i.e. with the fact that the variable $q^0$ is a cyclic one.}

 From this assumption it follows that
\be
    p_0= \frac{\pa \cL }{\pa \dq^0}
    :={\cP}_0 \bl (q^i(t),\dq^0(t),\dq^j(t),t \br )=\mbox{const},
\la{eq:1.2}
\ee
\enlargethispage{20pt}
where $i,j = 1,2,...,n$.

 Then, cf.~\cite{LL},
\begin{enumerate}
\item Equation (\ref{eq:1.2}) can be solved with respect to the variable
$\dq^0$ leaving us with a relation of the form
\[
\dq^0(t) =\phi \bl(p_0,q^i(t),\dq^j(t), t\br ),
\]
 where $p_0$ is an arbitrary, but fixed, value of the integration
constant. As a result, the variables $\bl(q^0(t),\dq^0(t)\br)$ can be
eliminated from the system of the $n+1$ original Lagrange equations. \item
The remaining $n$ differential equations for the variables $q^i(p_0,t)$
are again Euler-Lagrange equations of an action integral
 \be
   {\cW}_{p_0}[q^i] = \int\limits_{t_1}^{t_2} L_{p_0}
      \bl (q^i(t), \dq^j(t),t\br ) dt,  
\la{eq:1.3}
\ee
where $L_{p_0}$ is defined as
\be
  L_{p_0}\bl (q^i, \dq^j,t\br ) = 
\cR\bl(q^i,\phi(p_0,q^k,\dq^l,t),\dq^j,t\br),
\la{eq:1.4}
\ee
and where $\cR$ denotes the Routh function
\[
\cR(q^i,\dq^0,\dq^j,t) := \cL\bl (q^i,\dq^0,\dq^j, t \br) - \dq^0 p_0.
\]
\item After the Euler-Lagrange equations corresponding to the action
(\ref{eq:1.3}) have been solved for $q^i(p_0,t)$, one can find the 
function
$q^0(p_0,t)$ by solving the differential equation
\be
   \dq^0 = - \frac{\pa \cR_{p_0}}{\pa p_0}
           =\tilde  \phi (p_0,t), 
\la{eq:1.5}
\ee
where the function $\tilde \phi (p_0,t)$ is a solution of the
equation
\be
  {\cP}_0 \bl(q^i(p_0,t),\tilde \phi (p_0,t),\dq^j(p_0,t),t \br) =
             p_0 
\la{eq:1.6}
\ee
into which the now known functions $q^i(p_0,t)$ and $\dq^j(p_0,t)$
are substituted.
\end{enumerate}

\vspace{.4cm}
{\noindent\bf 3.~The Routh inverse procedure}

\vspace{.1cm}
{\noindent In order to determine the complete motion described by
$(q^0,q^i)$, the knowledge of a pair of functions $(L_{p_0},{\cP}_0)$, and
of a constant $p_0$ was necessary. A natural question now arises whether
this information is also sufficient to determine the functional form of
the original Lagrangian $\cL$ provided the triple $(L_{p_0},{\cP}_0,p_0)$
is known.}

The answer to the question just posed is positive, and the proof proceeds 
as follows.
\begin{enumerate}
\item Suppose that a function ${\cP}_0$ is given. Any 
Lagran\-gian $\cL(q^i,\dq^0,\dq^j,t)$ such that
\be
   \frac{\pa \cL }{\pa \dq^0}
    ={\cP}_0 \bl (q^i,\dq^0,\dq^j,t \br ) 
\la{eq:2.1}
\ee
is of the form
\be
      	 \cL(q^i,\dq^0,\dq^j,t)=
	                  I(q^i,\dq^0,\dq^j,t) + \Lambda(q^i,\dq^j,t),
\la{eq:2.2}
\ee
where 
\begin{displaymath}
I(q^i,\dq^0,\dq^j,t)=\int {\cP}_0(q^i,\dq^0,\dq^j,t)\, d \dq^0, 
\end{displaymath}
and $\Lambda$ is a quite arbitrary function of the arguments shown in
(\ref{eq:2.2}). 
\item The arbitrariness of $\Lambda$ is removed by the requirement 
that the Routh procedure, which starts from the assumption
\be  
   {\cP}_0 \bl(q^i(t),\dq^0(t),\dq^j(t),t \br)=p_0=\mbox{const},
\la{eq:2.3}
\ee
if applied to (\ref{eq:2.1}), lead to the now known Lagrangian 
$L_{p_0}(q^i,\dq^j,t)$. As a result, one obtains 
\be   
\Lambda(q^i,\dq^j,t)=    
 L_{p_0}(q^i,\dq^j)-I\bl(q^i,\varphi(p_0,q^k,\dq^l,t)\,\dq^j,t\br)
+\varphi(p_0,q^i,\dq^j,t) p_0,
\la{eq:2.4}
\ee 
where the function $\varphi$ is defined in an implicit way by the equation
\be
      {\cP}_0 \bl(q^i(t),\varphi,\dq^j(t),t \br)=p_0.
\la{eq:2.5}
\ee
 Of course, the value of the parameter $p_0$ in Eqs.~(\ref{eq:2.3}) and
(\ref{eq:2.5}) must agree with that entering the known Lagrangian
$L_{p_0}(q^i,\dq^j)$.
 \item The final functional form of $\cL$, obtained in consequence of 
substituting Eq.~(\ref{eq:2.4}) into Eq.~(\ref{eq:2.2}), is
 \be
\cL(q^i,\dq^0,\dq^j,t)= 
L_{p_0}(q^i,\dq^j,t)+\varphi(p_0,q^i,\dq^j,t)\,p_0 
+  \int\limits_{\varphi(p_0,q^i,\dq^j,t)}^{\dq^0}
  {\cP}_0 \bl (q^i,\kappa,\dq^j,t \br ) \, d\kappa.
\la{eq:2.6}
\ee
 Depending on the number of solutions for $\varphi$ admitted by
Eq.~(\ref{eq:2.5}), the solution (\ref{eq:2.6}) of the inverse problem may
not be a unique one.
 \end{enumerate}

\vspace{.4cm}
{\noindent\bf 4.~The Jacobi principle}

\vspace{.1cm}
{\noindent Let us consider now an action functional of the form
\be
     W[q] = \int\limits_{t_1}^{t_2} L
      \bl (q^i(t), \dq^j(t)\br ) dt. 
\la{eq:3.1}
\ee
The form above is equivalent to $\frac{\pa L}{\pa t}=0$ which implies the
energy conservation law $G(q^i,\dq^j) = E$, where
\be
 G(q^i, \dq^j) = \dq^i\,\frac{\pa L}{\pa \dq^i} - L  
\la{eq:3.2}
\ee
is the energy function, and $E$ is the energy constant.}

 In order to bring the action (\ref{eq:3.1}) to a form to which the Routh
formalism may be applied, a transformation of the parameter: $t\to \t$ is
performed, defined as $t=\te(\t)$, where $ \pt(\t)\neq 0$, and $\te$ is a
meanwhile unknown function. The action (\ref{eq:3.1}) transforms then into
 \be
   {\cW}[\te, x^i] = \int\limits_{\t_1}^{\t_2} \Lambda
      \bl (x^i(\t), \pt(\t), \px^j(\t)\br ) d\t,  
\la{eq:3.3}
\ee
where
\begin{eqnarray}
\hspace{-2cm}  \Lambda \bl (x^i(\t), \pt(\t), \px^j(\t)\br ) &\!\!=\!\!&
L\!\bl(x^i(\t), \frac{\px ^j(\t)}{\pt (\t)}\br )\,\pt(\t),\la{eq:3.4} \\
                         &   &   \nonumber \\
\mbox{and}\hspace{5cm} &   &   \nonumber \\
   x^i(\t) &\!\!:=\!\!& q^i\bl(\te(\t)\br), \la{eq:3.5}\\
 {\px}^{i}(\t)&\!\!:=\!\!& \dq^i\bl(\te(\t)\br)\,\pt(\t).\hspace{4.2cm} 
\la{eq:3.5'} 
\end{eqnarray}

The new Lagrangian $\Lambda$ determines a system of $n+1$ degrees of
freedom described by $n+1$ independent variables $(\te, x^i)$ being
functions of a parameter $\t$. (Notation like $\px =\frac{dx}{d\t}$ {\it
etc} is applied here).

The Lagrangian $\Lambda$ is a homogeneous function of degree one in the
variables $(\pt,\px^i)$. The appropriate variational principle determines
thus only $n$ independent differential equations of motion regardless of
the fact that the system is described by $n+1$ dynamical variables. The
Lagrangian $\Lambda$ does not explicitly depend on $\te$. Therefore, this
variable plays here the same role as $q^0$ did in the case of the
Lagrangian $\cL$ discussed before. Equation (\ref{eq:1.2}) reads now
 \begin{eqnarray}
    p_0&\!\!=\!\!&\frac{\pa \Lambda}{\pa \pt}\hspace{11cm} \nonumber \\
 &\!\!=\!\!&L\!\bl (x^i(\t), \frac{\px^j(\t)}{\pt(\t)}\br )
   -\frac{\px^k(\t)}{\pt(\t)}\,\frac{\pa L}{\pa \dq^k}\!
    \bl (x^i(\t),\frac{\px^j(\t)}{\pt(\t)}\br ) 
   =  -G\!\bl (x^i(\t),\frac{\px^j(\t)}{\pt(\t)}\br )\!.\hspace{0.6cm}
\la{eq:3.6} 
\end{eqnarray}
 Thus $\cP_0 =  - G(x^i,\frac{\px^j}{\pt})$, and $p_0=-E$. 
Therefore, we have to solve the equation
\be
    G\bl (x^i(\t),\frac{\px^j(\t)}{\pt}\br ) = E  
\la{eq:3.7}
\ee
with respect to $\pt$, prior to starting with the Routh formalism.

Writing the solution as
\be
     \pt(\t) =  \phi_E\bl ( x^i(\t), \px^j(\t)\br ),  \la{eq:3.8}
\ee
we are prepared to transform $\Lambda$ to a corresponding Routh function
which is denoted now by $L_E$,
\pagebreak
\begin{eqnarray}
 L_E(x^i, \px^j)&=&\Lambda \bl( x^i, \phi_E(x^j, \px^k), \px^l\br) - 
                    p_0\, \phi_E(x^i, \px^j)  \nonumber \\
&=&\bl[ L\bl(x^i, \frac{\px^j}{\phi_E(x^k,
         \px^l)}\br) + E \br] \phi_E(x^r, \px^s) \la{eq:3.9}\\
&=&\px^i \bl[ \frac{\pa L}{\pa \dq^i}\bl(x^k, \frac{\px^l}
                {\phi_E(x^r, \px^s)}\br)\br].  \la{eq:3.10} 
\end{eqnarray}

The Lagrangian $L_E$, for the first time derived by Jacobi, describes a
reduced dynamical system which resulted from eliminating the information
about the time evolution from the original system with the Lagrangian $L$.
In other words, the variables $q^i(t)$ which enter $L$, after the
corresponding equations of motion are solved, describe motions of the
system in $\Q^n$ which are curves in $\Q^n$ parametrized by the Newtonian
time $t$.  On the other hand, the variables $x^i$ that enter $L_E$
describe trajectories (i.e.~spatial paths) of the system; these
trajectories are only {\it loci} of points in $\Q^n$. As far the
computations that determine the form of the Lagrangian $L_E$ are
concerned, the expression (\ref{eq:3.9}) is, in my opinion, more suitable
for practical computations than the usually quoted expression
(\ref{eq:3.10}). It is worthwhile to note that the original Lagrangian $L$
provides information about the form of its energy function $G$, whereas
this piece of information is lost from the reduced Lagrangian $L_E$; from
Eq.~(\ref{eq:3.2}) it follows that its ``energy'' function identically
vanishes, i.e.~{\it no energy -- no time evolution}.

One can show, cf.~\cite{BJ}, that objects introduced in this section
have the following properties.
 \begin{enumerate}
 \item The function $\phi_E$ is homogeneous of degree one in the variables
$\px^i$, which means that the relation (\ref{eq:3.8}) is covariant with 
respect to reparametrizations $\t \to \t^\prime$.
 \item This in turn implies that also the Jacobi Lagrangian $L_E$ is
a homogeneous function of degree one in the variables $\px^i$. 
 \item The rank of the Hesse matrix of $L_E$ is equal to $n-1$.
\end{enumerate}

Points 2 and 3 mean that the Lagrange equations
\be
   \frac{\delta L_E}{\delta x^i} 
   := \frac{\pa L_E}{\pa x^i} - \frac{d}{d \t} 
     \bl(\frac{\pa L_E}{\pa \px^i}\br) = 0,  \la{eq:3.11}
\ee
together with appropriate initial conditions, can only determine 
trajectories in ${\Q}^n$ described by equations of the form
\be
F_K(q^1,\dots,\,q^n) = 0,\; \mbox{where} \; K=1,\dots,\,n-1,\la{eq:3.12}
\ee
or, usually under obvious additional assumptions, of the form $q^K= 
q^K(q^n)$.

To determine the complete motion $q^i=q^i(t)$ defined by the original
Lagrangian $L$, one has to add to the $n-1$ equations taken out from 
(\ref{eq:3.11}) the equation
\be
       G\bl ( q^i(t), \dq^j(t)\br )=E. \la{eq:3.13}
\ee
Thus, to determine the complete motion, one needs the triple $(L_E,G,E)$.
The pair $\bl (q^i(t), t\br )$ geometrically represents a world line in 
the space of states ${\Q}^n \times {\R}$ in which the unit taken along
the real axis $\R$ is equal to the unit of the Newtonian time $t$.

Remark. Equation (\ref{eq:3.13}) could as well be replaced by the 
equivalent 
equation
\[
                  \phi_E\bl ( q^i(t), \dq^j(t)\br )=1.
\]             

\vspace{.4cm}
{\noindent\bf 5.~The inverse Jacobi problem}

\vspace{.1cm}
{\noindent Let $L_h\bl(x^i(\t),\px^j(\t)\br)$ be a function homogeneous of
degree one in the variables $\px^i$. A variational principle with $L_h$
taken as the Lagrangian is only determining (non-parametrized) curves in a
${\Q}^n$.  The following questions can be asked here} 
\begin{enumerate}
 \item[i.] What data should be added to the knowledge of $L_h$, in order
to be able to lift the spatial paths in ${\Q}^n$ to motions $q^i=q^i(t)$
determined by a Lagrangian $L\bl (q^i(t),\dq^j(t)\br )$ such that the
given homogeneous Lagrangian $L_h$ is its Jacobi Lagrangian $L_E$
corresponding to $E$ taken as the energy constant?
 \item[ii.] What is the algorithm that enables us to determine $L$ in
terms of an arbitrarily given $L_h$ and what are the necessary additional
data that make the solution to the problem unique?
 \end {enumerate}

Problem of such a kind was formulated and solved in \cite{BJ} under
the name of {\it inverse Jacobi problem}. Now I would like to
present its solution.

All that said here so far suggests that a good candidate for the
additional data would be an arbitrarily assigned function $G(q^i,\dq^j)$
being the hoped-for energy function of the yet unknown Lagrangian $L$.

After introducing the velocity variable $v^i=\dq^i(t)$, relation (3.2)
turns into a partial differential equation
\be
v^1 \frac{\pa  L}{\pa v^1} + \ldots + v^n 
\frac{\pa  L}{\pa v^n}
-  L = G \la{eq:4.1}
\ee
for an unknown function $L(v^i)$. In Eq.~(\ref{eq:4.1}), $G=G(v^i)$ is
treated as a given function, and the dependence of $L$ and $G$ on $q^i$ is
here suppressed.

Applying the standard methods of integration of partial linear
differential equations, a general integral of (\ref{eq:4.1}) can be found 
to have the form 
\be
 L(q^i,v^j) = \sqrt{\mid g_{rs}v^r v^s\mid}\;
       I\!\!\left(q^i,\frac{v^j}{\sqrt{\mid g_{kl}v^k v^l\mid}},
                        \sqrt{\mid g_{pq}v^p v^q\mid}\right)
+ \Lambda(q^i,v^j),
\la{eq:4.2}
\ee
where $g_{ij}$ stands for the metric tensor in the manifold $\Q^n$ (in 
case such a tensor is absent, one may write down $g_{ij}=\delta_{ij}$), 
and where 
\be
   I(c^i,\rho):=\int \frac{G(c^i \rho)}{\rho^2}\,d\rho.
\la{eq:4.2'}
\ee

The function $\Lambda(q^i, v^j)$ in (\ref{eq:4.2}) is an arbitrary
integration function homogeneous of degree one in the variables $v^j$. The
equation (\ref{eq:4.2}) represents a general formula that determines a
class of Lagrangians $L$ describing a conservative dynamical system in
terms of an {\it a priori} assigned energy function $G$ of the system and
an arbitrary homogeneous Lagrangian $\Lambda$.

To solve the problem, we have to remove the arbitrariness of $\Lambda$ by
making use of the requirement that the given homogeneous Lagrangian
$L_h(x^i,\px^j)$ be the Jacobi Lagrangian corresponding to the Lagrangian
$L$ determined by Eq.~(\ref{eq:4.2}).
 
In order to be able to use the definition (\ref{eq:3.9}) of $L_E$, we have 
to find first the function $\phi_E$ by solving the equation
\be
    G\bl (x^i,\frac{\px^j}{\phi_E}\br ) = E.  \la{eq:4.3}
\ee

By using the requirement just mentioned, it is a quite simple technical
matter to find the function $\Lambda$ as a functional of $L_h$, $G$, and
$\phi_E$.

Substituting this functional into (\ref{eq:4.2}), one obtains the 
Lagrangian $L$ which solves the problem posed:
\begin{eqnarray}
 L(q^i, v^j)& = &\sqrt{\mid g_{ij}v^i v^j\mid}\,\left[
       I\!\!\left(q^i,\frac{v^j}{\sqrt{\mid g_{pq}v^p v^q\mid}},
\sqrt{\mid g_{rs}v^r v^s\mid}\right)\right. \nonumber \\
    &  &\bl. -I\!\!\left(q^i,\frac{v^j}{\sqrt{\mid g_{pq}v^p v^q\mid}}, 
\frac{\sqrt{\mid g_{rs}v^r v^s\mid}}{\phi_E(q^m,v^n)} \right)\right]  
           +   L_h(q^i, v^j) - E \phi_E(q^i, v^j). \hspace{20pt} 
  \la{eq:4.4}
\end{eqnarray}

\vspace{.4cm}
{\noindent\bf 6.~Geodesics in a Lorentzian manifold}

\vspace{.1cm}
{\noindent Let $g_{\al \bt}=g_{\al\bt}(\xi^\gamma)$, $\al ,\bt =
0,1,...,n$, be a Lorentzian metric in a local coordinate system
$\{\xi^\al\}$ in a manifold ${\M}^{n+1}$. The choice of its signature is
$+\, - \dots -$.\hspace{3pt} The geodesic lines $\xi^\al=\xi^\al(t)$ in
${\M}^{n+1}$, parametrized by an affine parameter $t$, are defined by the
action
 \be
{\cal W} = -{\textstyle\frac{1}{2}} \int\limits_{\t_1}^{\t_2} 
                      \,g_{\al\bt}\,u^\al\,u^\bt\, dt{,} 
\la{eq:5.1} 
\ee 
 where $u^\al=\frac{d\xi^\al}{dt^{\phantom{a}}}$. The action
(\ref{eq:5.1}) determines geodesics as {\it loci} of points in an
$n+2$-dimensional space $\R\times\M^{n+1}$, where $\R$ is the parameter
axis. The space $\R\times\M^{n+1}$ is here, unlike in Newtonian mechanics,
defined only locally over a geodesic line being just under consideration.
In the case of the action (\ref{eq:5.1}), let us denote its ``energy''
function by $\tilde{G}$. By making use of Eq.~(\ref{eq:3.2}), the function
$\tilde{G}$ can be found in the form
 \be
      \tilde{G}(\xi^\al,u^\bt)=-{\textstyle\frac{1}{2}}\,g_{\al\bt}\,
u^\al u^\bt. \la{eq:5.1'}
\ee 
 If one assigns now to the ``energy'' constant $C$ the value
\be
         C = -{\textstyle\frac{1}{2}}\,\vep\, m^2\,c^2, \la{eq:5.1a}
\ee
where $\vep=\pm 1$, and $m$ and $c$ are some constants, then by solving 
the Euler-Lagrange equations, with $m\neq0$, for $\vep=1$ one obtains 
timelike, and for $\vep=-1$ spacelike geodesics. The assumption $m=0$ is 
used here in case one wants to obtain a  null geodesic.

 Since the Lagrangian in the action (\ref{eq:5.1}) does not depend
explicitly on $t$, so it is possible here to perform the Jacobi
reduction. To this end, one must first solve Eq.~(\ref{eq:3.7}) in which
$G$ is replaced by $\tilde{G}$ from Eq.~(\ref{eq:5.1'}), and $E$ by $C$
defined in Eq.~(\ref{eq:5.1a}). Thus the solution (\ref{eq:3.8}) takes now
the form
  \be
  \pt(\t) = \phi_E(x^\al,\px^\bt)= \frac{1}{m\,c}\,
                          \sqrt{\vep\, g_{\al\bt}\,\px^\al\,\px^\bt},
 \la{eq:5.1''}
\ee
  where $x^\al(\t)=\xi^\al\bl(\te(\t)\br)$ and
$\px^\al=\frac{dx^\al}{d\t^{\phantom{a}}}$. Note that the Jacobi
reduction is not possible in the case of null geodesics.

Now with the aid of Eq.~(\ref{eq:3.9}), the Jacobi Lagrangian $L_C$ 
corresponding to the Lagrangian $\tilde{L}(\xi^\al,\px^\bt)$ of the action 
(\ref{eq:5.1}) can be easily found as
 \be
   L_C(x^\al,\px^\bt) = -\vep\, 
                        mc\,\sqrt{\vep\,g_{\al\bt}\,\px^\al\,\px^\bt}.
\la{eq:5.2} 
\ee 
 The Lagrangian $L_C(x^\al,\px^\bt)$ is homogeneous of degree one in
$\px^\al$, with all the consequences of this fact which were indicated
above. Thus, in case one would not like to introduce any additional
constraint condition, geodesics can be described analytically only by
equations of e.g.~the form $x^i=x^i(x^0)$, $i=1,\dots, n$. This means that
the geodesics are {\it loci} of points in the manifold $\M^{n+1}$, i.e.,
in this manifold,~they are world lines in the terminology used in the
theory of relativity.

\vspace{.4cm}
{\bf 7.~A Newtonian dynamical system}

\vspace{.1cm}
Let us consider in ${\Q}^{n}$ a system defined by the Lagrangian
\be
   \cL = {\textstyle\frac{1}{2}}\,e_{ij}\,v^i v^j + 
         {\textstyle\frac{e}{ c}}A_k v^k - V, \la{eq:6.1} 
\ee 
 where the notation is a standard one; $i,j = 1,...,n$. It is assumed that
the kinetic energy tensor $e_{ij}$, as well as the potentials $A_k$ and
$V$ are functions of only the coordinates $q^i$ in ${\Q}^{n}$, and they do
not depend explicitly on the time $t$. The system satisfies then the
energy conservation principle
 \be
 {\cal G} = {\textstyle\frac{1}{2}}\,e_{ij}\,v^i v^j + V = {\cal E}.  
\la{eq:6.2} 
\ee 
 If one wishes to apply to Eqs.~(\ref{eq:6.1}) and (\ref{eq:6.2})  the
Jacobi reduction procedure described in Sec.~4, one has to solve with
respect to $\pt$ first the algebraic equation
 \be
  \frac{\,e_{ij}\,\px^i \px^j}{2\,\pt}+ V = {\cal E} 
\la{eq:6.2'}
\ee
 corresponding in the present case to Eq.~(\ref{eq:3.7}), and next to
substitute into the equation which corresponds now to Eq.~(\ref{eq:3.9})  
the solution of Eq.~(\ref{eq:6.2'}), which is
 \be
 \pt =\phi_{\cal E}(x^k,\px^l)=\frac{\,e_{ij}\,\px^i \px^j}
                                    {2\,({\cal E}-V)}.
\la{eq:6.2''}
\ee
The outcome of all the operations just described is the
Jacobi Lagrangian   
\be
\cL_{\cal E} = \sqrt{2\,({\cal E} - V)\,e_{ij}\,\px^i\px^j} + 
              {\textstyle\frac{e}{c}}A_i\px^i 
\la{eq:6.3} 
\ee 
  of the system defined by the Lagrangian (\ref{eq:6.1}); for the notation
cf.~Eqs.~(\ref{eq:3.5}-\ref{eq:3.5'}). The Lagrangian (\ref{eq:6.3})
determines spatial paths in ${\Q}^{n}$, whereas the Lagrangian
(\ref{eq:6.1}) is defining motions in $\Q^n$ which could be looked upon as
world lines in ${\R}\times{\Q}^{n}$, where ${\R}$ is the Newtonian time
axis.}

\vspace{.4cm}
{\noindent\bf 8.~Geodesics in a stationary space-time}

\vspace{.1cm}
{\noindent Let
\be
   {\cal S} = -\vep\,mc\!\!\int\limits_{(q^0)_1}^{(q^0)_2}\!\! 
   \sqrt{\vep\left( g_{00} + 2\,g_{0k}\,\frac{dq^k}{dq^0}+
   g_{kl}\,\frac{dq^k}{dq^0}\frac{dq^l}{dq^0} \right) }\;dq^0
\la{eq:7.1}
\ee
 be an action for geodesics, $q^k=q^k(q^0)$, $k=1,...,n$, in a space
${\M}^{n+1}$ with coordinates $q^\al$ ($\al = 0,1,...,n$). It is assumed
that all $g_{\al \bt}$ do not depend explicitly on $q^0$. The minus sign
is standing here to assure a principle of the least action, as well as the
positive definiteness of energy.}

After replacing $q^0$ by $ t=\frac{q^0}{c^{\phantom{0}}}$, the Lagrangian
corresponding to (\ref{eq:7.1}) can be expressed as
 \be
   L(q^k,v^l)= -mc^2\,\vep\;\sqrt{\vep\left( g_{00} + 
2\,g_{0k}\,\frac{v^k}{c}+
   g_{kl}\,\frac{v^k v^l}{c^2} \right)}, \la{eq:7.2}
\ee
where $v^k=\dot q^k(t)$. Stationarity of ${\M}^{n+1}$ implies the
conservation of energy
\be
   G=mc^2\,\frac{g_{00}+g_{0k}\,{v^k/c}}{\sqrt{\vep\,\left( g_{00} + 
2\,g_{0k}\,{v^k/c}+ g_{kl}\,{(v^k v^l)/c^2} \br ) } } = E. \la{eq:7.3}
\ee

Let us apply now the Jacobi procedure presented in Sec.~4 to the
Lagrangian (\ref{eq:7.2}) taken together with its energy function
(\ref{eq:7.3}). It is a fairly straightforward matter to show that in
case the function $G$ is given by the expression (\ref{eq:7.3}), the
algebraic equation (\ref{eq:3.7}) on $\pt$ has a unique solution of the
form (\ref{eq:3.8}) in which for the function $\phi_E$ one must take
 \be
 \phi_E = - \frac{g_{0k}}{c\,g_{00}}\,\px^k + \frac{E}{c\,\sqrt{g_{00}}}
\,\sqrt{\frac{ \gamma_{ij}\,\px^i \px^j}{E^2/c^2 - m^2\, c^2 
\,\vep\,g_{00}}},
\la{eq:7.3'}
\ee   
where
\be
 \gamma_{ij}= -\left(g_{ij} - \frac{g_{0i}g_{0j}}{g_{00}}\right) 
\la{eq:7.5}
\ee
  is the so-called space metric tensor, cf.~\cite{LL}, and where the
notation introduced in Eqs.~(\ref{eq:3.5}) and (\ref{eq:3.5'}) applies.

With the aid of Eq.~(\ref{eq:3.9}), the corresponding Jacobi Lagrangian 
$L_E$ can now  be easily found as
\be
  L_E(x^i,\px^j) = \sqrt{\left(\frac{E^2}{c^2\,g_{00}} 
                   - m^2\,c^2\,\vep \right)\,\gamma_{ij}\,\px^i \px^j } 
                   -\frac{E\,g_{0k}}{c\,g_{00}}\,\px^k. 
\la{eq:7.4}
\ee	

The Lagrangian (\ref{eq:7.4}) determines spatial trajectories in
${\M}^{n}$ being a section of ${\M}^{n+1}$ with the hypersurface
$x^0=\mbox{const}\,$.

Let us note that due to the equivalence principle, the mass parameter $m$
that enters the Lagrangian (\ref{eq:7.2}), unlike the parameter $\vep$,
does not appear in the Euler-Lagrange equations of motion which follow
from this Lagrangian. These equations of motion admit however a whole
class of solutions for which 
 \be
     g_{00} + 2\,g_{0k}\,\frac{v^k}{c}+g_{kl}\,\frac{v^k v^l}{c^2}= 0, 
\la{eq:7.4'} 
\ee 
 i.e.~for which $L=0$. Thus these motions, represented by null geodesics
in $\M^{n+1}$, are not determined by an action principle based on the
action (\ref{eq:7.1}). Solving the energy conservation law (\ref{eq:7.3})
with respect to the square root of the expression standing on the
l.h.~side of Eq.~(\ref{eq:7.4'}), one can see that the square root tends
to zero for $m \to 0$ for the values of $\vep$ and $E$ kept fixed and
different from zero.  Therefore, vanishing of the expression in
(\ref{eq:7.4'}) can be considered to be equivalent to $\lim
m=0$.\footnote{Of course, a similar equivalence could have been obtained
by putting down $m=1$ and letting $\vep$ tend to zero.  The way accepted
in the article seems to be a more physical one. It demonstrates that a
null geodesic is a limiting case of either timelike or spacelike geodesics
which are selected by choosing either one of the two values of the
discrete parameter $\vep=\pm 1$, while the mass parameter accepts its
values from a continuous interval.}

From Eq.~(\ref{eq:7.4}) it follows that $L_E$ is a meaningful Jacobi
Lagrangian also for $m=0$. Therefore, despite the fact that the action
(\ref{eq:7.1}) does not work for null geodesics in
${\M}^{n+1}$, the corresponding Jacobi action based on the Lagrangian
$L_E$, given by Eq.~(\ref{eq:7.4}) for $m=0$, defines spatial paths in
${\M}^{n}$ of such geodesics in ${\M}^{n+1}$. In that case $L_E$ is
the Lagrangian of Fermat's principle for stationary space-times. This
principle, thought in a different theoretical framework, was already
discussed e.g.~in \cite{ACL}. 

For $m\neq 0$, the action principle based on the Lagrangian $L_E$ given by
(\ref{eq:7.4}) can be considered as being a generalization of Fermat's
principle for non-null geodesics in a stationary space time ${\M}^{n+1}$.
A Lagrangian of this kind, but only for static space times, was discussed
in \cite{Sz}, which unfortunately is a paper with many logical and
technical errors. In neither, however, of the papers just mentioned, the
true dynamical origin of the principles discussed there was revealed.

Let us finally observe that one can identify the manifolds ${\M}^{n}$
and ${\Q}^{n}$ of Sec.~7. This is implied by the fact that after making 
the identifications 
\begin{eqnarray}
   e_{ij}&=&\gamma_{ij}; \la{eq:7.6'} \\
    eA_i&=&-\frac{g_{0k}}{g_{00}}\,E; \la{eq:7.6} \\
   V &=& {\cal E} + {\textstyle\frac{1}{2}}\, m^2 c^2\vep - 
         \frac{E^2}{2\,g_{00}\,c^2}; 
\la{eq:7.6''}
\end{eqnarray}
and fixing the values of $m,\,e,\,{\cal E},\,E$, one can uniquely express
the quantities $g_{\al\bt}$ through $e_{ij},\,A_k,\,V$, or {\it vice 
versa}.
 
The identification of the spaces $\Q^n$ and $\M^n$ and the relations
(\ref{eq:7.6'})-(\ref{eq:7.6''}) demonstrate not only the equivalence of
the two dynamics defined, correspondingly, by $L_E$ and $\cL_{\cal E}$,
but they also reveal the existence of maps leading from e.g.~the dynamics
determined by $\cL$ to that by $L$ or the other way round, in accordance
with the diagrams
 \be
   {\cal L},\, (\cL,\,{\cal E})
    \buildrel {_{\mathrm{Jacobi}}} \over \longrightarrow 
   \cL_{\mathcal{E}},
   \begin{picture}(20,10)\put(5,0)
   {\scalebox{1.3}{$\exists$}}\put(1,-3){\mbox{$_{_{(G,\,E)}}$}}
   \end{picture}\; \mathcal{L}_{\mathcal{E}} + (G,\,E) 
   \buildrel{^{\rm inverse}_{\rm Jacobi}}\over\longrightarrow L,
   \la{eq:7.7}
\ee
and
\be
   L,\, (L,\, E)\buildrel{_{\rm Jacobi}}\over\longrightarrow L_E,
   \begin{picture}(20,10)\put(5,0)
   {\scalebox{1.3}{$\exists$}}\put(2,-3){\mbox
   {$_{_{({\cal G},\,{\cal E})}}$}}\end{picture}\;
    L_E +  ({\cal G},\,{\cal E})
   \buildrel{^{\rm inverse}_{\rm Jacobi}}\over\longrightarrow \cL,
   \la{eq:7.8}
\ee
 where the notation refers to objects that were already discussed in this 
article.

To demonstrate the way of how diagrams of this kind should be read, let us
explain it by taking the diagram (\ref{eq:7.7}) as an example. The
starting point here is the non-degenerate Lagrangian $\cL$ of the form
(\ref{eq:6.1})  which describes the motion of a system as a {\it locus} of
points in the $n+1$-dimensional space $\R \times \Q^n$. The knowledge of
$\cL$ uniquely determines, by means of Eq.~(\ref{eq:3.2}), the energy
function $\cal G$ given by (\ref{eq:6.2}). A choice that must be made
before the Jacobi reduction procedure is started is that of selecting a
value of the energy constant $\cal E$ in Eq.~(\ref{eq:6.2'}). Thus, to
start the Jacobi reduction, one has to select a pair $(\cL,{\cal E})$ in
$\R \times \Q^n$.  The outcome of the Jacobi procedure is a homogeneous
Lagrangian $\cL_{\cal E}$ which is made equivalent to the Lagrangian $L_E$
by means of Eqs.~(\ref{eq:7.6'})-(\ref{eq:7.6''}). Now, there exists a
pair $(G,E)$, consisting of a function $G(x^i, \px^j)$ and a value of a
constant $E$, such that when the piece of information encoded in the pair
is logically added to that encoded in the Lagrangian $\cL_{\cal E}$, one
obtains the starting point of an inverse Jacobi procedure that leads us to
the target Lagrangian $L(q^k,v^l)$ given by Eq.~(\ref{eq:7.2}). Of course,
for every $\cL_{\cal E}$ there is only one pair $(G,E)$ that allows us to
obtain the Lagrangian (\ref{eq:7.2}).

\vspace{.4cm}
 {\noindent\bf 9.~Geodesics  in a stationary space-time in an affine 
          parametrization}

\vspace{.1cm}
{\noindent The action (\ref{eq:7.1}) can be easily transformed to a 
homogeneous form.  
This may be achieved by introducing an additional dynamical variable
$q^0(\t)$ as a function of a new parameter $\t$. Its values are here
denoted by $x^0$, i.e.~$x^0=q^0(\t)$; and the remaining dynamical
variables are then transformed into $x^k=x^k(\t):=q^k\bl(q^0(\t)\br)$.  
After changing the integration variable $q^0\to\t$, $dq^0=\px^0\,d\t$, the
integral (\ref{eq:7.1}) takes the form
\be
   {\cal S}_h = -\vep\,mc\!\int\limits_{\t_1}^{\t_2}\!\! 
   \sqrt{\vep\, g_{\al \bt}\,\px^\al\px^\bt}\,d\t, 
\la{eq:8.1'}
\ee  
 where all $g_{\al \bt}$ in the integrand do not depend explicitly on
$x^0$. The two actions, given respectively by (\ref{eq:7.1}) and
(\ref{eq:8.1'}), determine the same {\it loci} of points in the space
$\M^{n+1}$ provided the function $x^0(\t)$ in (\ref{eq:8.1'}) is not a
fixed one, but it is treated like any other dynamical variable during the
variational procedure. This property of the action (\ref{eq:8.1'}) is due
to the fact that its Lagrangian is a homogeneous function of degree one in
the velocities $\px^\al$. Thus the two dynamics, defined by the actions
(\ref{eq:7.1}) and (\ref{eq:8.1'}) respectively, are mutually
equivalent.\footnote{In some old classical texts on differential geometry
the action (\ref{eq:8.1'}) is referred to as describing geodesics in an
{\it arbitrary} parametrization. This phrase is, however, slightly
confusing, for it is used to mean {\it in a not yet specified
parametrization}.}}

There are various methods of transforming the action (\ref{eq:8.1'}) into
another one which would give us solutions in a form of parametrized curves
in a suitably defined configuration space $\Q$ or, differently speaking,
solutions which are world lines in locally defined spaces $\R \times \Q$,
where $\R$ is the parameter axis. All such methods amount to adding new
information to that encoded in the action (\ref{eq:8.1'}). For instance,
one can substitute for the function $x^0(\t)$ in (\ref{eq:8.1'}) any
given, monotonous function $x^0= \tilde{x}^0(\t)$. This turns the
action (\ref{eq:8.1'}) into a non-homogeneous one which determines
parametrized curves $x^k = \xi^k(\t)$, where $\xi^k(\t):=
x^k\bl(\tilde{x}^0(\t)\br)$, in the space $\Q=\M^n$.

In this section, it is the inverse Jacobi procedure that is to be used.  
The Lagrangian of the homogeneous action (\ref{eq:8.1'}) is
 \be
          L_h(x^k,\px^\bt) = -\vep\, 
                        mc\,\sqrt{\vep\,g_{\al\bt}\,\px^\al \px^\bt},
\la{eq:8.1a} 
\ee 
 i.e.~it is formally equal to the Jacobi Lagrangian given by
Eq.~(\ref{eq:5.2}), but now it does not depend explicitly on $x^0$.  The
formal equality may indicate that the procedure we are going to use is, in
a sense, inverse to the reduction procedure discussed in Sec.~6. Thus,
seemingly, to trace the inverse procedure, it would be sufficient to read
the equations of Sec.~6 in a reverse order from Eq.~(\ref{eq:5.2}) to
(\ref{eq:5.1}). Although one could in this manner obtain a piece of
helpful information, yet the inverse Jacobi method is more than
this.  It is in a way a procedure of lifting dynamics of a certain type
from a configuration space $\Q$ to dynamics of a different type in the
space $\R\times\Q$, where $\R$ is the axis of a meanwhile unknown
parameter $t=\te(\t)$, which is implicitly introduced by a choice of an
``energy'' function. In principle, the choice of such a function is fairly
arbitrary. In practice, however, this choice may be a guess based on the
Jacobi reduction method applied to certain hoped-for target Lagrangians of
the inverse procedure.
 
In the present case, the starting homogeneous Lagrangian is $L_h$ given by 
Eq.~(\ref{eq:8.1a}), and the space $\Q=\M^{n+1}$. In accordance with 
Sec.~6, the ``energy'' function is chosen to be   
 \be
    \tilde{G}(\xi^k,u^\bt) := -{\textstyle\frac{1}{2}}\,g_{\al\bt\,}
                                u^\al u^\bt,     \la{eq:8.1}
\ee
 where $g_{\al\bt\,}=g_{\al\bt\,}(\xi^k)$,
$\xi^\al(t)=x^\al\bl((\te^{-1}(t)\br)$,
$u^\al=\frac{d\xi^\al}{dt^{\phantom{a}}}$, and $t=\te(\t)$, for
$\pt(\t)\neq 0$, is a new parameter. Also the choice of the ``energy''
constant $C$ is, in principle, arbitrary. In order, however, to obtain a
desired target Lagrangian, it is chosen, in accordance with
Eq.~(\ref{eq:5.1a}), as $C = -{\textstyle\frac{1}{2}}\,\vep\,m^2c^2$.  
The next step consists in solving Eq.~(\ref{eq:3.7}) adapted to the
present notation. Its solution is presented in Eq.~(\ref{eq:5.1''}). After
changing in the expression for $\phi_C(x^k,\px^\bt)$, given by
Eq.~(\ref{eq:5.1''}), the names of the variables from $(x^k,\px^\bt)$ to
$(\xi^k,u^\bt)$, we substitute this expression and that for the function
$\tilde{G}(\xi^k,u^\bt)$ given by Eq.~(\ref{eq:8.1}) into
Eq.~(\ref{eq:4.4}), to obtain the target Lagrangian in the form
 \be 
   \tilde{L}(\xi^k,u^\bt)  = -{\textstyle\frac{1}{2}}\, g_{\al\bt\,}
                                 u^\al u^\bt.   \la{eq:8.2}
 \ee  
 i.e.~a Lagrangian of the same form as that in the action (\ref{eq:5.1}),
but now the Lagrangian (\ref{eq:8.2}) does not depend explicitly on
$\xi^0$. The parameter $ t$ introduced by the choice of the
``energy'' function (\ref{eq:8.1}) is an affine one. The Lagrangian
(\ref{eq:8.2}) determines world lines in ${\R}\times{\M}^{n+1}$, where
${\R}$ stands for the $t$ axis. 

The Lagrangian (\ref{eq:8.2}) depends neither on $t$ nor on $\xi^0$. Its
independence of $t$ gave rise to the possibility of the Jacobi reduction
procedure which was performed in Sec.~6, and here it would restore the
starting Lagrangian $L_h$. Although the Lagrangian (\ref{eq:8.2}) is
independent of $\xi^0$, it does depend on
$u^0=\frac{d\xi^0}{dt^{\phantom{0}}}$, so $\xi^0$ is a typical cyclic
variable, and the existence of such a variable enables us to apply the
Routh reduction procedure to the Lagrangian $\tilde{L}$ as well.

For the sake of this procedure, we replace in the Lagrangian $\tilde{L}$
the names of the variables $u^\al$ with $v^\al$ and write down
Eq.~(\ref{eq:8.2}) in a way that explicitly exposes the dependence of
$\tilde{L}$ on the variable $v^0$:
 \be
  \tilde{L}(\xi^k,v^\bt) = -{\textstyle\frac{1}{2}}\,
        \left(g_{00}(v^0)^2 + 2\,g_{0k}v^0v^k + g_{kl}v^k v^l\br),
\la{eq:8.3} 
\ee 
 In order to eliminate from the Lagrangian (\ref{eq:8.3}) the variables
$(x^0,v^0)$, we have to compute the quantity ${\cal P}_0$ defined in
Eq.~(\ref{eq:1.2}) for the case considered now. We have
 \be
 \frac{\pa \tilde{L}}{\pa v^0}= {\cal P}_0 := -g_{00}v^0-g_{0k}v^k=p_0,
\la{eq:8.3'}
\ee
 and the solution for $v^0$ of the last equation above is
 \be
     v^0 = \phi\bl(p_0,\xi^i(t),v^j(t)\br ):=
            -\frac{p_0}{g_{00}} - \frac{g_{0k}v^k}{g_{00}}.
\la{eq:8.3''}
\ee
 The Lagrangian (\ref{eq:1.4}) equals the Routh function
$\cR\bl(\xi^i,\phi(p_0,\xi^k,v^l),v^j,t\br)$ and takes now the form
 \be
   \cL_{p_0}(\xi^k,v^l) = {\textstyle\frac{1}{2}}\,\gamma_{kl}v^k v^l +
               p_0\,\frac{g_{ok}}{g_{00}}\,v^k + 
	       \frac{p_0^2}{2\,mg_{00}}.
\la{eq:8.4} 
\ee                                             
 The Lagrangian above is of the same type as that defined by
Eq.~(\ref{eq:6.1}). Therefore, we can identify the configuration space of
dynamics defined by the Lagrangian (\ref{eq:8.4}) with the configuration
space $\R\times\Q^n$ of the dynamics discussed in Sec.~7. Comparing in the
two Lagrangians, $\cL$ and $\cL_{p_0}$ respectively, the coefficients at
the same powers of $v^k$, we obtain
 \begin{eqnarray}
   e_{ij}&=&\gamma_{ij}; \la{eq:8.4a} \\
    eA_i&=&p_0 c\;\frac{g_{0k}}{g_{00}} \la{eq:8.4b} \\
   V &=& -\frac{{p_0}^2}{2\, g_{00}}+ \mbox{const}.  \la{eq:8.4c}
\end{eqnarray} 
 Comparing next the two sets of relations, represented respectively by
Eqs.~(\ref{eq:7.6'})-(\ref{eq:7.6''}) and by
Eqs.~(\ref{eq:8.4a})-(\ref{eq:8.4c}), we see that Eqs.~(\ref{eq:7.6'}) and
(\ref{eq:8.4a}) are identical, and Eq.~(\ref{eq:7.6}) and (\ref{eq:8.4b})
can be made identical by assuming that $p_0\,c = -E$. Then
Eq.~(\ref{eq:8.4c}) turns into
 \be
     V = -\frac{E^2}{2\, g_{00}\,c^2}+ \mbox{const},
\la{eq:8.4d}
\ee
 which demonstrates that the two dynamics, defined respectively by $\cL$
and $\cL_{p_0}$, are fully equivalent.

The content of this section is a generalization of a result by Eisenhart
\cite{E} who has shown that the trajectories of a general holonomic
conservative system of $n$ degrees of freedom in classical dynamics can be
put into correspondence with geodesics of a suitable Riemannian manifold
$\cal S$, where $\dim {\cal S}=n+1$. In \cite{E}, however, no use of
methods of analytical dynamics was made, in particular of those concerning
cyclic variables, but instead only tedious transformations of the
underlying ODE were performed.

Another reason which enables us to consider the result just obtained as a
more general one than that of Eisenhart is that it permits one to prolong
the sequence of mappings shown in the diagram (\ref{eq:7.7}) by a new
sequence presented in the following diagram
 \be
   L \buildrel{^{\mathrm{reparame-}}_{\rm trization}}\over
   \longrightarrow L_h,
   \begin{picture}(20,10)\put(5,0)
   {\scalebox{1.3}{$\exists$}}\put(1,-3){\mbox
   {$_{_{(\tilde{G},\,C)}}$}}\end{picture}\; 
   L_h  + (\tilde{G},\,C) 
   \buildrel{^{\rm inverse}_{\rm Jacobi}}\over\longrightarrow 
   \tilde{L},\,(\tilde{L},p_0) \buildrel{_{\rm Routh}}\over
   \longrightarrow \mathcal{L}_{p_0}\equiv \mathcal{L}. 
   \la{eq:8.5}
\ee 
 The complete sequence, made by joining the sequences (\ref{eq:8.5}) and 
(\ref{eq:7.7}) one after the other, forms a closed loop. In an analogous 
way, with the help of the algorithms presented in this article, 
one can also prove the validity of the following sequence of mappings
 \be
   {\cal L},\begin{picture}(20,10)\put(7,0)
   {\scalebox{1.3}{$\exists$}}\put(1,-3){\mbox
   {$_{_{({\cal P}_0,\,p_0)}}$}}\end{picture}\;\;\;  
    {\cal L} + ({\cal P}_0,\,p_0)\,
   \buildrel{^{\rm inverse}_{\rm Routh}}\over\longrightarrow 
   \tilde{L},\,(\tilde{L},\,C) \buildrel{_{\rm Jacobi}}\over
   \longrightarrow\, L_h \equiv L. 
\la{eq:8.6} 
\ee
 The composition \{(\ref{eq:7.8}), (\ref{eq:8.6})\} of the corresponding
sequences forms again a loop of mappings which passes exactly through the
same dynamics as the previous loop, but the other way round.

Thus the two loops, \{(\ref{eq:7.7}), (\ref{eq:8.5})\} and
\{(\ref{eq:7.8}), (\ref{eq:8.6})\} taken together, define a commuting
diagram of mappings between all the dynamics discussed above. In terms of
pairs consisting of Lagrangians and spaces of states\footnote{In the
terminology introduced by Synge \cite{Sy}, a {\it space of states} of
a dynamical system is the space in which the motions determined by the
dynamics are represented by curves being {\it loci} of points.} of
corresponding dynamics, the diagram may be shown as  
   \begin{eqnarray}
(\tilde{L},\,{\R}\!\!&\!\!\!\times{\M}^{n+1})\, \longleftrightarrow
\hspace{10pt} 
(L_h\!\!\!\!\!&,{\M}^{n+1}) \equiv (L,\,{\M}^{n+1})   \nonumber \\
    \updownarrow &              &\updownarrow  \la{eq:8.7} \\
({\cal L},\,{\R}\!\!&\!\!\! 
\times\,{\Q}^{n})\hspace{15pt}\longleftrightarrow
\hspace{10pt}  
({\cal L}_{\cal E}\!\!\!\!\!&,\,{\Q}^{n}) \equiv 
(L_E,\,{\M}^{n}).\nonumber 
\end{eqnarray}     

{\noindent In this diagram the names of the procedures which labelled the
corresponding mapping arrows, as well as other details concerning the
definitions of mappings are suppressed, but they may be easily recovered
by means of the diagrams (\ref{eq:7.7}), (\ref{eq:7.8}), (\ref{eq:8.5}),
and (\ref{eq:8.6}).}

It is rather remarkable that the seemingly arbitrary constant in
Eq.~(\ref{eq:8.4c}) can be easily determined. This follows from the fact
that the two dynamics, $(\tilde{L},\,{\R}\times{\M}^{n+1})$ and $({\cal
L},\,{\R}\times{\Q}^{n})$, are invariant under translations of respective
parameters $t$ in the two configuration spaces. And this fact was not
exploited yet.  The invariance induces in the space
${\R}\times\,{\M}^{n+1}$ the conservation law: $\,\tilde{G}(\xi^k, u^\bt)
= -{\textstyle \frac{1}{2}}\,\vep\,m^2c^2$, in accordance with
Eqs.~(\ref{eq:8.4c}) and (\ref{eq:5.1a}). In order to project this
conservation law on the space ${\R}\times\,{\Q}^{n}$, one must replace 
in
it the variables $u^\al$ with $v^\al$, eliminate from it the variable
$v^0$, and make use of Eq.~(\ref{eq:8.3''}), replacing $p_0\,c$ by $-E$.
After all this is done, one obtains
 \be
-{\textstyle\frac{1}{2}}\,\gamma_{ij}v^iv^j+\frac{E^2}{2\,g_{00}\,c^2}=
 {\textstyle\frac{1}{2}}\,\vep\, m^2c^2.
\la{eq:8.8}
\ee
 On the other hand, the same invariance of the dynamics $({\cal
L},\,{\R}\times{\Q}^{n})$ induces the conservation law (\ref{eq:6.2}).
Upon making use of the relation (\ref{eq:8.4a}), and eliminating from
Eq.~(\ref{eq:6.2}) the potential $V$ by means of Eq.~(\ref{eq:8.4d}),
one transforms Eq.~(\ref{eq:6.2}) into
 \be
 {\textstyle\frac{1}{2}}\,\gamma_{ij}v^iv^j - 
                   \frac{E^2}{2g_{00}c^2}={\cal E} - \mbox{const}\,{.}        
\la{eq:8.9}
\ee
 Eliminating now the kinetic term from Eqs.~(\ref{eq:8.8}) and
(\ref{eq:8.9}), one finds that
 \be
 \mbox{const} = {\cal E} + \frac{1}{2}\,\vep\, m^2c^2,
\la{eq:8.10}
\ee
 which shows that the relations (\ref{eq:7.6''}) and (\ref{eq:8.4d}) agree
with each other.

The last result indicates that the mappings from the diagram
(\ref{eq:8.7}) preserve various features of the three types of geodesics,
labelled by the values of $\vep$ and $m$.

\vspace{.4cm}
{\noindent\bf Acknowledgments}

\vspace{.1cm}
{\noindent This work was supported in part by the Polish Research 
Programme KBN, grant no.~2 P03B 127 24.}


\begin{thebibliography}{9}
\bibitem{LL}
           L.~D.~Landau and E.~M.~Lifshitz, {\it Mechanics},  Addisson 
           - Wesley, Reading, MA, 1960.
\bibitem{BJ} 
           S.~L.~Ba\.za\'nski, P.~Jaranowski, {\it The inverse Jacobi
           problem}, J.~Phys.~A: Math.~Gen., {\bf 27}, 3321 (1994).
\bibitem{B}
           S.~L.~Ba\.za\'nski, {\it The Jacobi Variational Principle 
           revisited}, in: Classical and Quantum Integrability,
           Banach Center Publications, Vol.~59, eds.~J.~Grabowski,
           G.~Marmo, and P.~Urba\'nski, Polish Academy of Sciences, 
           Inst.~of Mathematics, Warszawa, 2003.
\bibitem{ACL} 
           M.~A.~Abramowicz, B.~Carter, and J.~P.~Lasota, {\it Optical
           Reference Geometry for stationary and static dynamics},
	   Gen.~Rel.~Grav., {\bf 20}, 1173 (1988).
\bibitem{E} 
           L.~P.~Eisenhart, {\it Dynamical trajectories and 
           geodesics}, Ann.~Math., {\bf 30}, 591 (1929).
\bibitem{Sz} 
            M.~Szyd{\l}owski,
            {\it Desingularization of Jacobi metrics
            and chaos in general relativity}, J.~Math.~Phys., {\bf 40}, 
	    3519 (1999).
\bibitem{Sy}
           J.~L.~Synge, {\it Classical Dynamics}, in: Handbuch der
           Physik, Encyclopedia of Physics, Vol.~III/1, 
           ed.~S~Fl\"uge/Marburg, Springer, Berlin, 1960.

\end{thebibliography}
\end{document}